\documentstyle[amsfonts,12pt]{article}

\def\half{\textstyle{\frac{1}{2}}}

\def\E{{\rm I}\hskip-.2em{\rm E}}
\def\ra{\rightarrow}
\def\tint{{\textstyle\int}}
\def\hg{{\hat g}}
\def\hp{{\hat\pi}}

\def\s{\hskip.08em}
\def\l{\lambda}

\def\a{\alpha}
\def\b{\begin{eqnarray*}}     
\def\e{\end{eqnarray*}}       
\def\bn{\begin{eqnarray}}     
\def\en{\end{eqnarray}}       
\def\<{\langle}
\def\>{\rangle}

\def\no{\nonumber}

\def\dnx{d^{n}\!x}

\def\{{\lbrace}
\def\}{\rbrace}
\begin{document}
\title{Attractions of Affine Quantum Gravity\footnote{Presented at the 
Sixth Intl.~Symp.~on the Frontiers of Fund.~Physics, Udine, Italy, Sept., 
2004}}
\author{John R. Klauder
\footnote{Electronic mail: klauder@phys.ufl.edu}\\
Departments of Physics and Mathematics\\
University of Florida\\
Gainesville, FL  32611}
\date{}     
\maketitle
\begin{abstract}
All attempts to quantize gravity face several difficult problems. 
Among these problems are: (i) metric positivity (positivity of the spatial
distance between distinct points), (ii) the presence of anomalies 
(partial second-class nature of the quantum constraints), and (iii) 
perturbative nonrenormalizability (the need for infinitely many 
distinct counterterms). In this report, a relatively nontechnical 
discussion is presented about how the program of affine quantum gravity 
proposes to deal with these problems.
\end{abstract}
\subsection*{Introduction and Survey}
The program of affine quantum gravity differs from that of string theory 
or loop quantum gravity: specifically, it differs in the 
insistence on a spatial 
metric tensor that is strictly positive definite; in the simultaneous 
and uniform treatment of both first- and second-class operator constraints; in 
dealing with nonperturbative renormalizability; and in maintaining 
a close connection with the motivating classical (Einstein) gravity 
theory. A suitable realization of these principles is most readily 
presented within a formalism that is, for the most part, generally 
unfamiliar to many readers. The purpose of this paper is to provide
a relatively simple introduction to several concepts used to study 
quantum gravity from this new perspective.

\subsubsection*{ Reproducing Kernel Hilbert Spaces} 
This important idea can be
readily explained. Let $|l\>\in{\frak H}$, for all $l\in{\cal L}$, 
denote a set of states (chosen to be normalized, but that is not a 
requirement) that span the separable Hilbert space $\frak H$ of interest. In 
addition we assume that (when it is finite dimensional) the space 
$\cal L$ is locally equivalent to a Euclidean space, and that the
states $|l\>$ are continuously labeled by the (multi-dimensional)
labels $l$. We will refer to the set of states $\{|l\>\}$ as {\it coherent
states}.

Since the coherent states span $\frak H$, it follows that two elements
of a dense set of states may be given by
 \b &&|\psi\>=\Sigma_{j=1}^J \a_j\s|l_j\>\;,\hskip1cm J<\infty\;,  \\
  &&|\phi\>=\Sigma_{k=1}^K \beta_k\s|l_{(k)}\>\;,\hskip1cm K<\infty\;.  \e
As {\it functional representatives} of these abstract vectors we introduce
  \b && \psi(l)\equiv \<l|\psi\> = \Sigma_{j=1}^J \a_j\s\<l|l_j\>\;,\\
  &&\phi(l)\equiv \<l|\phi\> = \Sigma_{k=1}^K \beta_k\s\<l|l_{(k)}\>\;. \e
Finally, as the {\it inner product} for these two elements we choose
  \b (\psi,\phi)\equiv \Sigma_{j,k=1}^{J,K}\a^*_j\s\beta_k\s\<l_j|l_{(k)}\>=
\<\psi|\phi\>\;. \e
This pre-Hilbert space is completed by adding the limit points of Cauchy 
sequences. The result is a functional representation composed entirely of 
continuous functions representing the separable Hilbert space $\frak H$. 

Such spaces are called reproducing kernel Hilbert spaces because if 
$J=1$ and $\a_1=1$, then it follows that $\psi(l) =\<l|l_1\>$ and so
  \b (\psi,\phi)=\Sigma_{k=1}^{K}\s\beta_k\s\<l_1|l_{(k)}\>=\phi(l_1) \;,\e
a result which ``reproduces'' the vector $\phi(l)$. Thus the coherent state
overlap $\<l|l'\>$ serves as the ``reproducing kernel'' for this space. 

Observe that all properties of this representation are determined by the
jointly continuous coherent state overlap function $\<l|l'\>$. Indeed,
{\it any} continuous function of two (sets of) variables $K(l;l')$ serves to
define a reproducing kernel Hilbert space provided $K$ satisfies the condition
  $$ \Sigma_{j,k=1}^{J,J}\a^*_j\s\a_k\s K(l_j;l_k) \ge 0 $$
for all possible complex choices of $\{\a_j\}$ and finite $J$. 

\subsubsection*{Metric Positivity} 
Distinct points in a space-like 3-dimensional manifold have a positive 
separation distance. For a small coordinate separation $dx^a\not\equiv0$, 
that distance, as usual, is given by $ds^2=g_{ab}(x)\s
dx^a\s dx^b>0$. We require that the associated quantum operator $\hg_{ab}(x)$ 
also satisfy metric positivity such that $\hg_{ab}(x)\s dx^a\s dx^b>0$ in
the sense of operators for all nonvanishing $dx^a$. Moreover, we insist 
that $\hg_{ab}(x)$ becomes self adjoint when smeared with a suitable
real test function. In canonical quantization one chooses the canonical (ADM) momentum $\pi^{ab}(x)$ as the field to promote to an operator, $\hp^{ab}(x)$. 
However, since the momentum acts to translate the metric, such a choice is
inconsistent with the preservation of metric positivity. Instead, it is
appropriate to choose the mixed valence momentum field $\pi^a_c(x)\equiv 
\pi^{ab}(x)\s g_{bc}(x)$ to promote to an operator $\hp^a_c(x)$. This choice
is dictated by the relation
  \b e^{i\tint \gamma^a_b(y)\s\hp^b_a(y)\s d^3\!y}\,\hg_{cd}(x)\,
e^{-i\tint \gamma^a_b(y)\s\hp^b_a(y)\s d^3\!y}=(\s e^{\gamma(x)/2}\s)^e_c\;
\hg_{ef}(x)\,(\s e^{\gamma(x)/2}\s)^f_d\;,  \e
a relation that manifestly preserves metric positivity.

The full set of kinematical commutation relations is given by \cite{klagra}
 \b 
  &&\hskip.2cm[\hp^a_b(x),\,\hp^c_d(y)]=\half\s i\s[\s\delta^c_b\hp^a_d(x)
-\delta^a_d\hp^c_b(x)\s]\,\delta(x,y)\;,\no\\
  &&\hskip.1cm[\hg_{ab}(x),\,\hp^c_d(y)]=\half\s i\s[\s\delta^c_a\hg_{bd}(x)
+\delta^c_b\hg_{ad}(x)\s]\,\delta(x,y)\;,\label{afc}\\
&&[\hg_{ab}(x),\,\hg_{cd}(y)]=0\;.   \no \e
These are the so-called affine commutation relations appropriate to the affine 
fields $\hp^a_c(x)$ and $\hg_{ab}(x)$, both of which may be taken as self 
adjoint when smeared with real test functions. These commutation relations 
provide a realization of the group ${\rm IGL}(3,{\mathbb R})$, and 
as such they are more
in the spirit of a current algebra than traditional canonical commutation
relations. 

It is important to add that by choosing $\hp^a_c(x)$ as the partner field to
go with $\hg_{ab}(x)$, it follows that the momentum field $\hp^{ab}(x)$ 
does {\it not} make an operator when smeared but only a form.

\subsubsection*{Quantization of Constraints}
There are several schemes in common usage to quantize canonical systems
with constraints. Traditionally, these schemes treat first- and second-class 
constraints differently. Gravity is not a traditional gauge theory since
the set of classical constraints form an {\it open} first class system, 
which means that the Poisson brackets among the constraints have the form 
of a Lie algebra except that the structure constants are actually structure 
functions depending on the canonical variables. On quantization, these 
structure functions become operators that do not commute with the 
constraints, and as a consequence, the quantum constraints are partially 
second class in character. As noted above this usually entails a separate 
procedure for their analysis.

However, the recently introduced {\it projection operator method} \cite{proj} 
to incorporate quantum constraints treats first- and second-class 
constraints in an identical fashion and thereby it seems ideal to apply 
to gravity. Here, we content ourselves with a sketch of how this procedure 
is applied to simple systems.

We start by assuming that $\phi_\a(p,q)=0$, $\a=1,..., A$, represent a set
of real classical constraints for some multi-dimensional system. We choose 
some quantization procedure and identify $\{\Phi_\a(P,Q)\}$ as a set of 
self-adjoint operators representing the constraints. Ideally, following 
Dirac, we would identify the physical Hilbert space ${\frak H}_{phys}
\subset{\frak H}$ as composed of vectors
$|\psi_{phys}\>$ that have the property that
  \b \Phi_\a(P,Q)\s|\psi_{phys}\>=0  \e
for all $\a$. Consistency of this procedure requires that (i) 
$\<\psi_{phys}|\psi_{phys}\><\infty$, and (ii) $[\Phi_\a(P,Q),
\Phi_\beta(P,Q)]\s|\psi_{phys}\>=0$. Unfortunately, for certain constrained 
systems, either one or both of these
consistency conditions is violated. In that case it is useful to propose 
another scheme. 

One alternative procedure, known as the projection operator method, 
involves a 
projection operator 
  \b  \E=\E(\Sigma_\a\s\Phi^2_\a(P,Q)\le\delta(\hbar)^2)\;, \e
an expression which means that 
  \b 0\le\E\s\Sigma_\a\Phi^2_\a(P,Q)\s\E\le\delta(\hbar)^2 I \;.  \e
In these expressions, $\delta(\hbar)$ denotes a small, positive cutoff,
generally dependent on $\hbar$, that can be reduced to a suitable level.
In this approach the (regularized) physical Hilbert space is taken as
${\frak H}_{phys}=\E\s{\frak H}$. 

It is pedagogically useful to illustrate this procedure with three simple 
examples: \vskip.2cm

(1) Let $\Phi_\a=J_1,\,J_2,\,J_3$ denote the generators of SU(2), and the 
desired physical Hilbert space satisfies $J_k\s|\psi_{phys}\>=0$ for $k=1,2,3$.
We can secure the physical Hilbert space of interest by choosing
   \b  \E(\Sigma_k J_k^2\le \half\hbar^2)\;.  \e
This example represents a first-class constrained system. \vskip.2cm

(2) Let $\Phi_\a =P,\,Q$, a pair of canonical operators. In this case we choose
  \b \E(P^2+Q^2\le\hbar) \;,  \e
which projects onto states $|\psi_{phys}\>$ that satisfy 
$(Q+iP)\s|\psi_{phys}\>=0$. This example represents a second-class 
constrained system. \vskip.2cm

(3) Let $\Phi_\a=Q$, a single operator with zero in the continuous spectrum
and for which $Q\s|\psi_{phys}\>=0$ has no normalizable solution. here we 
choose
   \b \E(Q^2\le\delta^2)\;,  \e
with no $\hbar$ dependence necessary. As $\delta\ra0$, this projection 
operator passes strongly (hence weakly) to the zero operator. To overcome 
this fact, we {\it rescale} 
the projection operator and take a suitable limit as $\delta$ goes to zero.
As one example, we introduce coherent states 
  \b |p,q\>\equiv e^{-iqP}\s e^{ipQ}\s|0\> \;, \e
and we consider
\b \<\!\<p'',q''|p',q'\>\!\>\equiv \lim_{\delta\ra0}\,\frac{\<p'',q''|
\s\E(Q^2\le\delta^2)\s|p',q'\>}{\<0|\s\E(Q^2\le\delta^2)\s|0\>}\;. \e
The resultant expression forms a suitable reduced reproducing kernel which 
can be used to
characterize the physical Hilbert space as a reproducing kernel Hilbert space.

If $P$ and $Q$ form an irreducible pair, and for the sake of illustration we
choose $|0\>$ as a normalized solution of $(Q+iP)\s|0\>=0$, i.e., $|0\>$ is
the oscillator ground state, then
  \b \<\!\<p'',q''|p',q'\>\!\>=e^{-\half(q''^2+q'^2)}\;,  \e
a reproducing kernel which characterizes a {\it one}-dimensional Hilbert space.
Different choices of the fiducial vector $|0\>$ may lead to different
functional representatives, but they nevertheless still describe 
one-dimensional Hilbert spaces.

It is noteworthy that path integral expressions exist that directly generate 
matrix elements of any desired projection operator. For example, staying with
elementary examples, coherent state path integrals that generate expressions 
such as $\<p'',q''|\s\E\s|p',q'\>$ may formally be written as conventional 
phase-space path integrals save for one change, namely, the choice of the 
integration measure for the Lagrange multipliers \cite{proj}.

How these general ideas may be applied to quantum gravity can be found in 
\cite{klagra}.

\subsubsection*{Perturbative Nonrenormalizability}
One of the most challenging aspects of conventional approaches to quantum
gravity is its perturbative nonrenormalizability. Divergences can be 
regularized by the introduction of cutoffs, as usual, and then counterterms 
developed on the basis of perturbation theory can be identified and 
included in the formalism. For renormalizable theories there are only a finite 
number of distinct
types of counterterms, while for nonrenormalizable theories -- such as 
gravity -- an infinite set of qualitatively distinct counterterms is mandated 
by perturbation theory. It is no wonder that the morass created by 
renormalized 
perturbation theory has driven many workers to alternative approaches such 
as string theory. On the other hand, perhaps we are deceiving ourselves; 
could it be that perturbatively suggested counterterms to nonrenormalizable 
models are in fact {\it irrelevant}? This heretical viewpoint is indeed 
suggested by the 
{\it hard-core picture} of nonrenormalizable interactions which 
we now outline \cite{book}.

To present the essential ideas as simply as possible let us initially 
examine certain singular potentials in quantum mechanics. In particular, 
consider the Euclidean-space path integral for a free particle in the
presence of an additional potential $V(x)\ge0$. In symbols, let us study
 \b W_\l\equiv {\cal N}\int_{x(0)=x'}^{x(T)=x''}\, e^{-\half\tint 
{\dot x}(t)^2\s dt-\l\tint V(x(t))\s dt}\;{\cal D}x \;. \label{e37}\e
As $\l\ra0^+$, it appears self evident that $W_\l$  passes to the
expression
 \b W_0\equiv{\cal N}\int_{x(0)=x'}^{x(T)=x''}\, e^{-\half\tint{\dot x}^2
(t)\s dt}\;{\cal D}x=\frac{1}{\sqrt{2\pi T}}\,e^{-(x''-x')^2/2T}  \e
appropriate to a free particle. Whatever the analytic dependence of 
$W_\l-W_0$ for small $\l$ (e.g., $ O(\l),\,O(\l^{1/3}),\,O(e^{-1/\l})$, etc.), 
it is tacitly
assumed that as $\l\ra0^+$, $W_\l\ra W_0$, i.e., that $W_\l$ is {\it 
continuously connected} to $W_0$. However, this limiting behavior is {\it not} 
always true.

Consider the example $V(x)=x^{-4}$. In this case the singularity at $x=0$ 
is so strong that the contribution from all paths that reach or cross the 
origin is {\it completely suppressed} since $\tint x(t)^{-4}\s dt=\infty$ 
for such
paths, no matter how small $\l>0$ is chosen. As a consequence, as $\l\ra0^+$
for $V(x)=x^{-4}$, it follows that
\b \lim_{\l\ra0^+}W_\l=W'_0\equiv\frac{\theta(x''x')}{\sqrt{2\pi T}}
\bigg[\s e^{-(x''-x')^2/2T}-e^{-(x''+x')^2/2T}\s\bigg]\;.  \e
Stated otherwise, when $V(x)=x^{-4}$, $W_\l$ is {\it decidedly not} 
continuously connected to the
free theory $W_0$, but is instead continuously connected to an alternative
theory -- called a pseudofree theory -- that accounts for 
the {\it hard-core} 
effects of the interaction. The interacting theory may well possess a 
perturbation expansion about the pseudofree theory (to which it is 
continuously
connected), but the interacting theory will {\it not} possess any 
perturbation expansion about the free theory (to which it is not even 
continuously 
connected). 

Let us next pass to scalar field theory and the Euclidean-space functional 
integral 
 \b S_\l(h)\equiv {\cal N}\int \exp\{{\tint} {h\phi\s\dnx}-\half
\tint[(\nabla\phi)^2+m^2\phi^2]\s\dnx-\l\tint\phi^4\s\dnx\}\;{\cal D}\phi  \e
appropriate to the $\phi^4_n$ model in $n$  spacetime dimensions.  We recall
for such expressions that there is a Sobolev-type inequality to the effect
that
  \b  \{\tint\phi(x)^4\s\dnx\}^{1/2}\le K\tint[(\nabla\phi(x))^2+m^2\phi(x)^2]
\s\dnx \e
holds for {\it finite} $K$ (e.g., $K=4/3$) whenever $n\le 4$, but which 
{\it fails} to hold (i.e., $K=\infty$) whenever $n\ge5$. Thus for 
nonrenormalizable
interactions $\phi^4_n$, for which $n\ge5$, it follows that there are fields
$\phi$ for which the free action is finite while the interaction action is 
infinite. Just as in the elementary example, there is no reason to believe
that counterterms suggested by a regularized perturbation analysis (the
underlying premise of which is to maintain a continuous connection with
the free theory!) should have any relevance in defining the pseudofree
theory $S'_0(h)$. 

It is noteworthy that proposals have been advanced to define $S'_0(h)$ and 
thereby to develop a meaningful and nontrivial theory of nonrenormalizable
scalar fields \cite{lmp}. Monte Carlo studies of such proposals are
currently under way.

Lastly we observe that gravity is also a theory for which the free action 
(limited to quadratic terms) does not dominate the interaction action 
(remaining terms),
and consequently gravity would seem to be a candidate theory to be 
understood on the basis of a hard-core interaction, which, when regularized, 
leads to its perturbatively nonrenormalizable behavior.
As plausible as this scenario seems, it will involve a considerable effort
to establish it convincingly.

\end{document}